\documentclass[12pt]{article}
\pdfoutput=1

\usepackage[utf8]{inputenc}
\usepackage[english]{babel}
\usepackage{comment}

\usepackage{slashed}
\usepackage{amssymb,amsmath,amsfonts}
\usepackage{graphicx}
\usepackage{fullpage}
\usepackage{physics}
\usepackage{cite}
\usepackage{appendix}
\usepackage{cmap}
\usepackage{csquotes}

\title{
\vspace{-2cm} 
\begin{flushright}
{\normalsize INR-TH-2023-021}
\end{flushright}
\vspace{0.5cm} 
Shower formation constraints on cubic Lorentz Invariance Violation parameters in quantum electrodynamics}
\author{Petr Satunin$^{1,2}$\thanks{{\bf e-mail}: satunin@ms2.inr.ac.ru}, Andrey Sharofeev$^{2,1}$\thanks{{\bf e-mail}: sharofeev@ms2.inr.ac.ru}
\vspace{.2cm}\\
\normalsize\it  $^1$ Institute for Nuclear Research of the Russian Academy of Sciences, \\  
\normalsize \it  60th October Anniversary Prospect, 7a, 117312  Moscow, Russia
\\
\normalsize\it $^2$ Department of Particle Physics and Cosmology, Physics Faculty, \\ \normalsize \it Moscow State University, \\ 
\normalsize \it Leninskiye Gory, 119991 Moscow, Russia}
\date{}

\begin{document}
\maketitle
\begin{abstract}
We present shower formation constraints on the Lorentz Invariance Violation (LIV) energy scale for photons with  cubic dispersion relation from recent gamma ray observations in $100$ TeV -- PeV energy range by LHAASO observatory. We assume Myers-Pospelov effective field theory framework, and calculate the suppression for the Bethe-Heitler process which is mainly responsible for the formation of photon-initiated atmosphere showers. Comparing the high-energy photon events with the suppressed flux predictions we obtain $95\%$ CL constraints on the LIV energy scale. 
The obtained shower constraint  $E_{\rm LIV} \gtrsim \mathcal{O} \left( 10^{20} \mbox{ GeV}\right)$ is significantly weaker than existing birefringence constraints but is independent.

\end{abstract}


\section{Introduction}

The formalism of quantum field theory 
based on some fundamental symmetries including Lorentz Invariance (LI) as a cornerstone. However, possible deviations from LI
may appear in several scenarios for quantum gravity (see~\cite{Addazi:2021xuf} for a review, \cite{Liberati_2013, mattingly2005modern} and reference therein). Traditionally, in the effective field theory (EFT) approach one can gather all possible LIV operators for the Standard Model fields into so-called Standard Model Extension (SME)~\cite{Colladay:1998fq}. While low-order LIV terms are usually tested with precise laboratory experiments~\cite{Kostelecky:2008ts}, high-order 
terms seem irrelevant at laboratory energies but may be significant at higher energies. 

High-energy cosmic rays are the best tool for these tests since its energy in the laboratory frame is larger than those achievable at ground-based accelerators. Hypothetical LIV can influence different processes relevant to production, propagation and detection of these cosmic rays. A pure separate channel of the SME is the quantum electrodynamics (QED) sector. The number of parameters in the QED sector reduces significantly, so it can be deeply studied. The key but not only characteristic of LIV of this type is the modification of dispersion relations,
\begin{equation}
    E^2=p^2  \pm \frac{p^3}{E_{\rm{LIV},1}} \pm \frac{p^4}{E_{\rm{LIV},2}^2} + ...
\end{equation}
Thus, extra terms proportional to high powers of energy suppressed by large energy scale appear. 





The most consistent consideration of the case leading to LIV can be investigated through the construction of EFTs, as it was noted above. From such theories, for example, a non-standard form of dispersion relations can immediately follow. The advantage of considering such theories allows one to use the whole formalism of quantum field theory, thus obtaining most consistently, for example, cross sections of processes. The only EFT leading  to the cubic momentum violation of LI in the QED 
is the Myers-Pospelov model~\cite{PhysRevLett.90.211601} which is considered in the current article. 

LIV induces some processes forbidden in LI and modifies thresholds of  existing processes~\cite{Coleman_1997, Coleman_1999, Jacobson_2006}. For example, some LIV allows channels of photon decay which are forbidden in LI, such as $\gamma \to e^+e^-$ and $\gamma \to 3\gamma$. Being allowed, these processes eliminate photons of given energy from the flux.
Additionally, LIV can influence air shower formation. Particularly, the cross-section of the most probable first reaction in air showers --- the Bethe-Heitler process  --- modified in case of LIV which leads to deeper air showers~\cite{Vankov:2002gt}. Detection of normal air showers sets constraints on LIV parameters. 

The paper is organized as follows. In Sec.~\ref{sec:Myers-Pospelov model -- an EFT for cubic LIV}, we discuss the Myers-Pospelov EFT and its basis properties.
In Sec.~\ref{sec:Pair production on a nuclei and Air shower formation}, we discuss the air shower formation by very-high-energy photons and the following constraints to the observed flux. In Sec.~\ref{sec:Observational bounds -- LHAASO}, we estimate observational bounds at the LHAASO experiment for three astrophysical sources. The Sec.~\ref{sec:Discussion} is devoted to conclusions.

\section{Myers-Pospelov model -- an EFT for cubic LIV}
\label{sec:Myers-Pospelov model -- an EFT for cubic LIV}

The full EFT Lagrangian which leads to cubic LIV in  particle dispersion relations has been written by Myers and Pospelov~\cite{PhysRevLett.90.211601}. The part of the Myers-Pospelov action related separately to photons reads, 
\begin{equation}
    \mathcal{L}_\gamma = -\frac{1}{4}F_{\mu\nu}F_{\mu\nu} + \frac{\xi}{M_{\rm{Pl}}} n^\mu F_{\mu \nu} n \cdot \partial \left( n_\sigma \Tilde{F}^{\sigma \nu}\right).
    \label{MPLagrangian}
\end{equation}
The parameter $\xi$ is a dimensionless constant, $M_{\rm{Pl}}$ is the Planck mass. 
The four--vector $n^\mu$ is a purely time-like vector, $n^\mu = (1,0,0,0)$. Note that CPT parity is violated as well.



The equations of motion in the model~(\ref{MPLagrangian}) reads, 
\begin{equation}
    \Box{A^\tau} = \frac{\xi}{M_{\rm{Pl}}} n_\sigma \epsilon^{\sigma \tau \rho \nu} \left(n \cdot \partial \right)^2 F_{\rho \nu},
    \label{EquationofMotion}
\end{equation}
The two transverse polarizations of the photon, determined by the equations of motion~(\ref{EquationofMotion}), have the following dispersion relations:
\begin{equation}
 \omega^2 = k^2 \pm \frac{k^3}{E_{\rm LIV}}, 
    \label{DispersionRelation}
\end{equation}
where the sign plus (minus) related to left (right) polarized photon, and we introduce the typical energy scale for LIV,
\begin{equation}
    E_{\rm LIV} \equiv \frac{M_{\rm Pl}}{2 \xi}.
\end{equation}
The main effect of modified dispersion relation~(\ref{DispersionRelation}) is that the group velocity of traveling photons depends on its energy and polarization, $v = 1 \pm k/E_{\rm LIV}$, where we recall left polarization related to the sign \enquote{$+$} as \enquote{superluminal} polarization, right polarization related to the sign \enquote{$-$} --- as \enquote{subluminal} one.

The clue difference with the LIV QED, quartic on spatial momentum~\cite{Rubtsov:2012kb} is that in quartic theory both  polarizations have the same dispersion relation with specific sign, \enquote{$+$} or \enquote{$-$}. So, the \enquote{super-} and \enquote{subluminal} photons are related to independent sectors of model parameters in that model, and the constraints on these sectors are set independently. Contrary, in our cubic model for fixed LIV mass scale one has both super-- and subluminal photons simultaneously. Two phenomena based on modified velocities and important in bounding $E_{\rm LIV}$ are time delays and vacuum birefringence.

\paragraph{Time delays.}

Due to the energy dependence for the photon velocity, photons traveling from cosmologically distant astrophysical sources would acquire additional dispersion. The absence of such phenomena constrain $E_{\rm LIV}$. The  actual constraint came from  GRB090510, $E_{\rm LIV} > 9.3\cdot 10^{19}$ GeV~\cite{Vasileiou:2013vra}.

\paragraph{Vacuum birefringence.} 

This phenomenon manifests itself as the rotation of the polarization plane for a photon beam. Even small rotation, being accumulated  over vast cosmological distances, may be  potentially detectable, see Sec. 5.2.1 of~\cite{Addazi:2021xuf} for details. 
The most stringent constraint on the mass scale came from  polarized GRB061122~\cite{Gotz:2013dwa},  $E_{\rm LIV} \sim \mathcal{O} \left( 10^{34} \mbox{ GeV} \right)$.

This constraint is  sufficiently trans-Planckian and may close the Quantum Gravity motivation for these models. Nevertheless, independent consideration of cubic LIV from threshold processes may also be interesting. Any contradictions, if they appear, may lead to a breakdown of the EFT approach.

\paragraph{Modification of perturbative processes} The cross-sections of several perturbative processes including photons in the model~(\ref{MPLagrangian}) get modified compared to the standard case. Moreover, super- and subluminal photons may lead to different outcomes. Superluminal photons have an excess of energy compared to the momentum and can release it in any allowed decay process. Subluminal photons, in turn, have a lack of energy any interaction channel of the subluminal photon gets suppressed.

\paragraph{Photon decay}
The simplest perturbative process is the photon decay. Superluminal photon polarization undergoes photon decay to an electron-positron pair for a photon with energy larger than the threshold one, $E^{\rm th}_\gamma = \sqrt[3]{4 m^2 E_{\rm LIV}}$, $m$ is the mass of an electron. Once kinematically allowed, the decay occurs almost immediately. Since, the detection of even one superluminal photon with energy $E_\gamma$ sets the bound,
\begin{equation}
 \label{DecayBound}
    E_{\rm LIV} > \frac{E_\gamma^3}{4 m^2}.
\end{equation}
The corresponding bound for superluminal photons are well-studied in the literature~\cite{Coleman_1997,Coleman_1999,Jacobson:2002hd,Jacobson_2006}.  Another possible decay channel for the superluminal polarization is the photon splitting~\cite{Gelmini:2005gy,Astapov:2019xmt}.


Let us turn to what does mean the constraint on super(or sub)luminal photons in the context of the Myers-Pospelov model. At high energies of the photon  the photon polarization cannot be easily measured in the process of the detection. Thus, if we detect a photon of a given energy, one cannot immediately write the bound~(\ref{DecayBound}) but first study the outcomes if it was a photon of a subluminal polarization.



Two processes whose modifications are important in case of subluminal photon dispersion relation are pair production by a high energy photon in a Coulomb field (Bethe-Heitler process), and pair production on soft background photons.

\section{Pair production on nuclei and air shower formation}
\label{sec:Pair production on a nuclei and Air shower formation}

The first process suppressed for subluminal  photon
is the Bethe-Heitler process --- photon decay to an electron-positron pair in the Coulomb field of a nuclei. This process is responsible for the first interaction of astrophysical photons in the atmosphere. 
The suppression of the Bethe-Heitler process in a subluminal LIV scenario was firstly estimated by~\cite{Vankov:2002gt}.  


Consequently, in such scenarios, atmospheric showers initiated by photons would penetrate more deeply into the atmosphere compared to the conventional scenario. These exceptionally deep showers may remain undetected during experiments. Therefore, the forecast for subluminal LIV resembles that of the superluminal case, both resulting in the suppression of photon flux, particularly for the highest-energy photons. Hence, in the subluminal scenario, photon-induced air showers would commence at greater atmospheric depths compared to the standard case.

In the event that the initial photon interaction's depth, represented as $X_0$, within the atmosphere surpasses the overall atmospheric depth, denoted as $X_{\rm{atm}}$, the development of the shower will be impeded, resulting in the event remaining undetected.

The probability for a photon to produce pair in the atmosphere reads,

\begin{equation}
    P = \int\limits_0^{X_{\rm{atm}}} \dd{X_0} \frac{e^{-X_0/\expval{X_0}_{\rm{LIV}}}}{\expval{X_0}_{\rm{LIV}}} = 1 - e^{-X_{\rm{atm}}/\expval{X_0}_{\rm{LIV}}},
    \label{Probability}
\end{equation}
where the mean depth of the interaction $\expval{X_0}_{LIV}$ in the LIV case can be expressed via the mean depth $\expval{X_0}_{LI} = 57 \mbox{ g cm}^{-2}$ in the LI case and the ratio of the cross sections $\sigma^{\rm{LIV}}$ and $\sigma^{\rm{LI}}$ as follows,
\begin{equation}
    \expval{X_0}_{\rm{LIV}} = \frac{\sigma^{\rm{LI}}}{\sigma^{\rm{LIV}}} \expval{X_0}_{\rm{LI}}.
\end{equation}

The ratio of cross sections will be used as follows,

\begin{equation}
    \frac{\sigma^{\rm{LIV}}}{\sigma^{\rm{LI}}} \simeq 1.7 \cdot \frac{m^2 E_{\rm{LIV}}}{E_\gamma^3} \log \frac{E_\gamma^3}{2 m^2 E_{\rm{LIV}}}.
    \label{Ratio of cross sections}
\end{equation}

We see that the cross section in LIV case decreases with energy as $E_\gamma^{-3} \log E_\gamma$ under the condition that $E_{\rm{LIV}} \ll E_\gamma^3/2m^2$.

Using~(\ref{Ratio of cross sections}) we can transform~(\ref{Probability}) in the following way,

\begin{equation}
    P(E_\gamma, E_{\rm LIV}) = 1 - \exp( -1.7 \cdot \frac{X_{\rm{atm}}}{57 \mbox{ g cm}^{-2}} \frac{m^2 E_{\rm{LIV}}}{E_\gamma^3}  \log \frac{E_\gamma^3}{2 m^2 E_{\rm{LIV}}}).
    \label{Final probability}
\end{equation}


The prediction for the observed flux can be formulated as follows:

\begin{equation}
\label{FLUX}
    \left(\dv{\Phi}{E} \right)_{\rm{LIV}} = P(E_\gamma, E_{\rm LIV}) \times \eval{\dv{\Phi}{E}}_{\rm{source}}.
\end{equation}

Here, we use the suppression factor denoted as $P(E_\gamma, E_{\rm LIV})$, which is defined in equation~(\ref{Final probability}), to assess the potential for particle shower formation. The hypothesis regarding a specific value of $P(E_\gamma, E_{\rm LIV})$ can be tested with experimental data. It is important to note that conducting these tests requires selecting a model that describes the photon flux from the source. To obtain the most reliable constraints on the LIV parameter $E_{\rm LIV}$, when considering multiple source models, we choose the model that predicts the highest photon flux.

\subsection{Pair production on EBL and CMB}
High energy photons, traveling in soft photon backgrounds, undergo the pair production process, $\gamma\gamma_b \to e^+ e^-$, which results in the attenuation of the photon flux $e^{-\tau}$ . This process goes if photon energy exceed the threshold~\cite{Galaverni:2007tq},

\begin{equation}
    E_{\rm{th}} = \frac{m^2}{\omega_b} \mp \frac{1}{4} \frac{k^2}{E_{\rm{LIV}}},
\end{equation}
where the sign \enquote{$+$} goes from the subluminal case, the sign \enquote{$-$} goes from the superluminal case.


The cross-section of the process above the threshold remains relatively insensitive to LIV~\cite{PhysRevD.86.085012}. A significant consequence of this process is the reduction in the flux of high-energy photons, typically in the TeV energy range, originating from extragalactic sources as they interact with extragalactic background light (EBL)~\cite{Biteau_2015}. In the case of subluminal dispersion relation, the pair production threshold shifts to higher energies of background photons, which is associated with lower EBL density, leading to less attenuation in the extragalactic photon spectrum~\cite{Kifune_1999, Stecker:2001vb}. The absence of this phenomenon places constraints on the viability of the subluminal dispersion relation~\cite{sym12081232, PierreAuger:2021tog}. When considering galactic gamma rays within the energy range of 1 to 100 TeV, one can safely disregard the pair production effect due to the relatively short distances within the galaxy compared to the mean free path of pair production in the presence of the infrared background.

The detection of galactic photons with energies below a PeV by Tibet-AS$\gamma$ and LHAASO has brought about changes in the threshold for pair production, shifting it into the range of cosmic microwave background (CMB) energy levels. Notably, the density of CMB surpasses that of extragalactic background light (EBL) by over two orders of magnitude. This implies that the mean free path for a photon with an energy of approximately $\sim \mbox{ PeV}$, involved in CMB-induced pair production, is around $\sim 10 \mbox { kpc}$, a length scale comparable to galactic scales~\cite{PhysRev.155.1408, Vernetto:2016alq}.

The attenuation of galactic gamma-ray flux can be expressed through the attenuation factor denoted as $P_{\rm CMB}$, given by the exponential function 

\begin{equation}
    P_{\rm CMB}(L_{\rm source},E_\gamma,E_{\rm LIV}) = \exp(-L_{\rm source} / L_{\rm MFP}(E_\gamma, E_{\rm LIV})),
    \label{ProbabilityAttenuation}
\end{equation}
where $L_{\rm MFP}$ represents the mean free path of photons. In the scenario of subluminal LIV, the threshold for pair production is elevated above the energy corresponding to the peak of the CMB spectrum, causing the attenuation effect to vanish. Consequently, in the subluminal LIV framework, the modification of pair production results in an increase in the observed gamma-ray flux, while the shower suppression effect leads to a reduction.

Nevertheless, it is worth noting that for the spectra observed in references~\cite{cao2021ultrahigh, Amenomori:2019rjd}, the factor $P_{\rm CMB}$ cannot fall below $0.5$. Therefore, although  the impact of modified pair production on CMB is secondary in comparison to the shower suppression effect it should be definitely taken into account by dividing eq.~(\ref{FLUX}) on~(\ref{ProbabilityAttenuation}),
\begin{equation}
    \left(\dv{\Phi}{E} \right)_{\rm{LIV}} = \frac{  P(E_\gamma, E_{\rm LIV})}{P_{\rm CMB}(L_{\rm source},E_\gamma,E_{\rm LIV})} \times \eval{\dv{\Phi}{E}}_{\rm{source}}.
\end{equation}
This formula is to test the hypothesis of a certain $E_{\rm LIV}$



\section{Observational bounds -- LHAASO}
\label{sec:Observational bounds -- LHAASO}




The Large High Altitude Air Shower Observatory (LHAASO), located at $4,410$ meters above sea level in Daocheng, China, began its first-stage operations in 2019. It will have sensitivity to gamma rays up to $\mbox{TeV}-\mbox{PeV}$ range and cosmic rays in the $\mbox{TeV}$ to $10^3 \mbox{ PeV}$ range. LHAASO has recently reported spectrum of 12 sources  in our galaxy with energy up to $1.4 \mbox{ PeV}$~\cite{cao2021ultrahigh}, and Crab later  based on the observation of the highest-energy photons ever detected, up to $1.4 \mbox{ PeV}$, and later the photons from Crab~\cite{LHAASO:2021cbz} with implications for the LIV purposes~\cite{cyranoski2019china, cao2021ultrahigh, Li:2021tcw, Chen:2021hen, Satunin:2021vfx}. Particularly, the 3 gamma-ray sources with energy, which were provided by LHAASO, especially, J2226+6057 and J1908+0621~\cite{cao2021ultrahigh}, Crab Nebula~\cite{LHAASO:2021cbz} were considered in this work to constrain $E_{\rm LIV}$ with shower formation.

The maximum zenith angle at which LHAASO events occur is 50 degrees~\cite{cao2021ultrahigh}. So, one can compute the maximum atmospheric depth for inclined showers at the LHAASO altitude, $X_{\rm atm} = 890 \mbox{ g } \mbox{cm}^{-2}$.

We introduce the following criterion for the selection of bins in the case of J2226+6057, J1908+0621, and Crab Nebula events. On one hand, we consider only those energy bins that exhibit a significance greater than $2\sigma$, therefore, we omit the last bin for J1908+0621 and the two last bins for J2226+6057. On the other hand, we do not use energy bins in cases where the variation between the LI and LIV scenarios is negligible. 
\begin{figure}[h]
    \centering
    \includegraphics[width=0.49\textwidth]{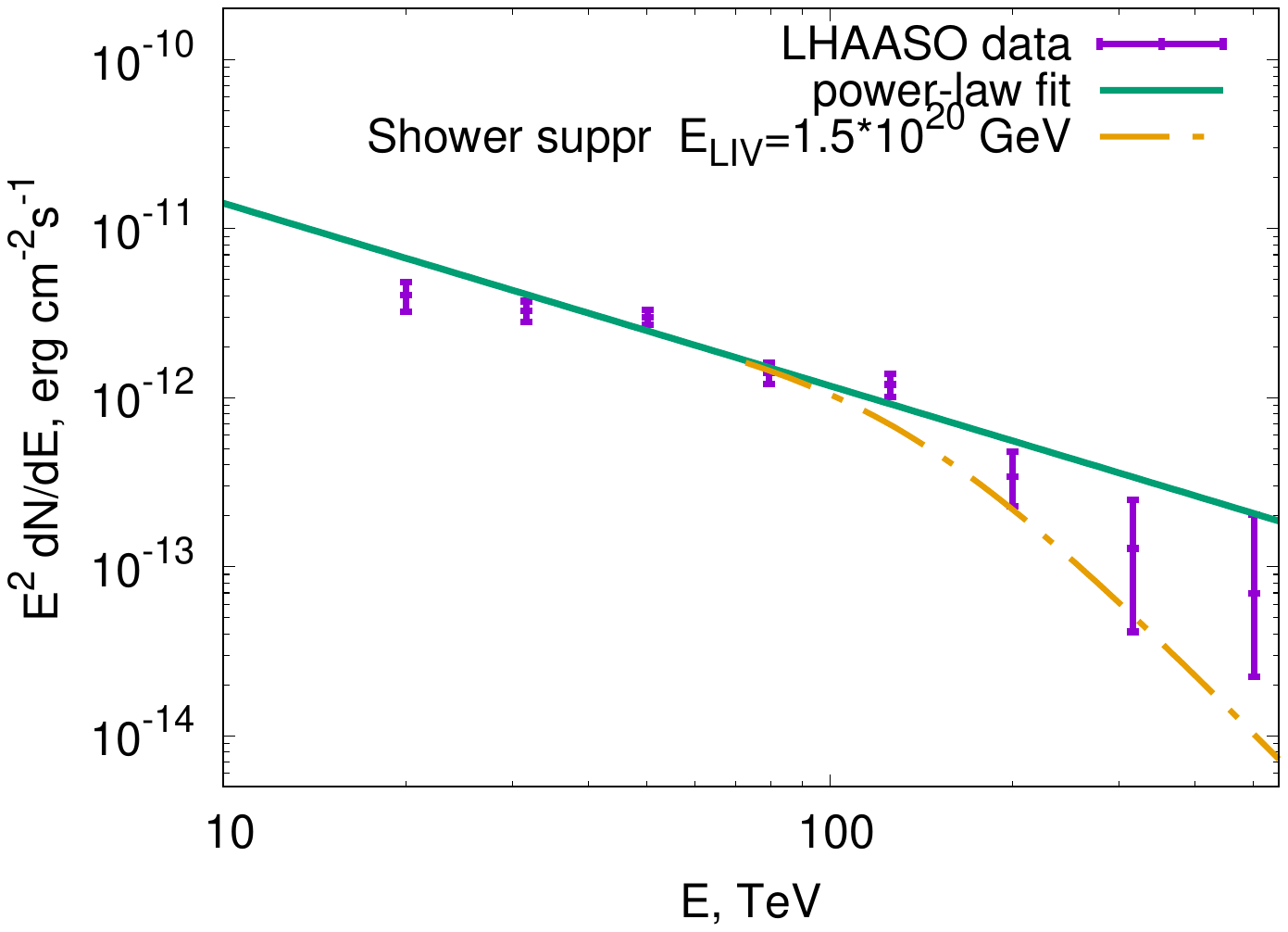} 
    \includegraphics[width=0.49\textwidth]{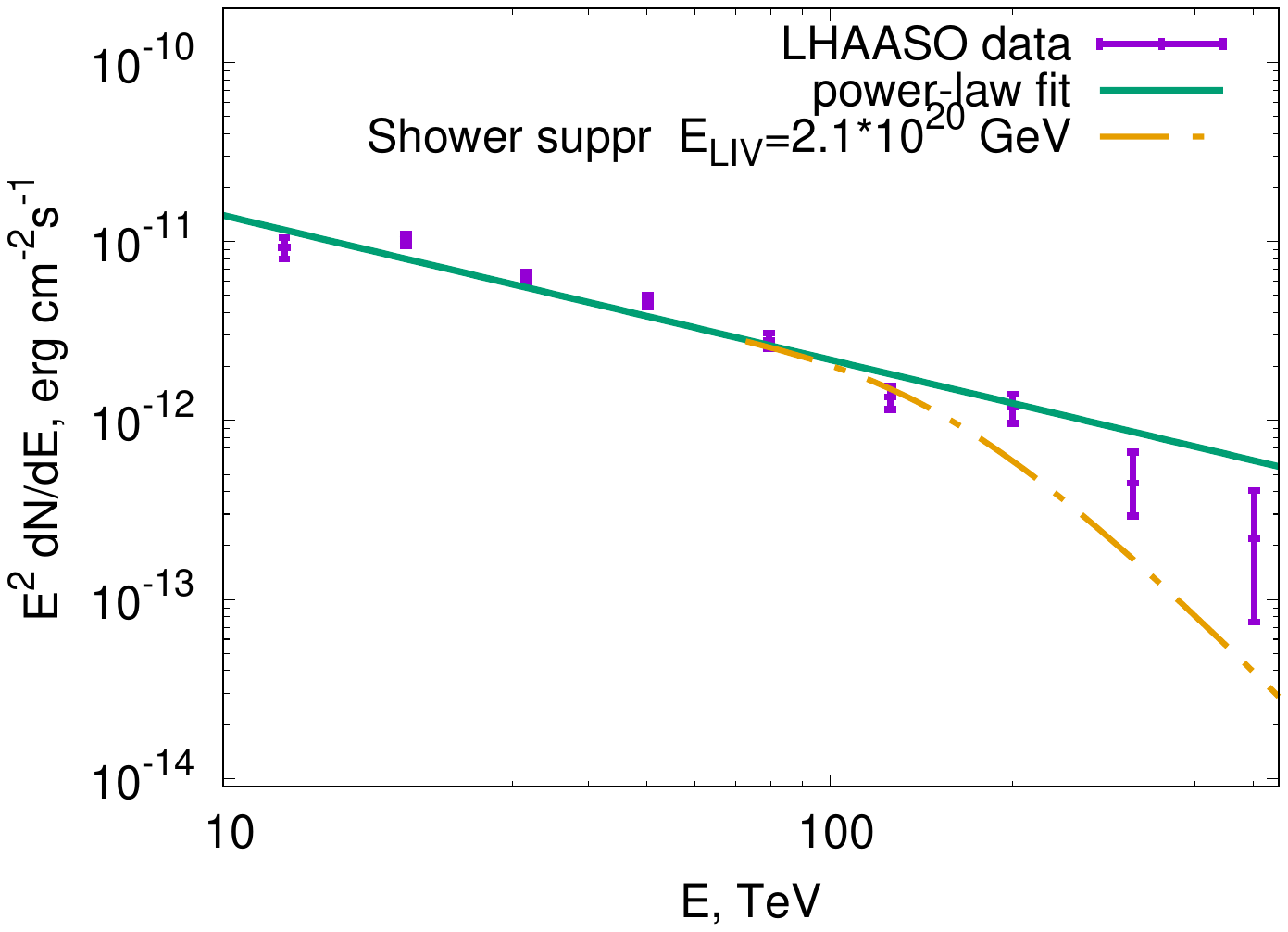} \\
    \caption{The galactic source spectra measured by LHAASO (data points), the power-law fit for the spectra (green solid curve), the predicted flux under  subluminal LIV (yellow dash-dotted curve). Left panel: J2226+6057, Right panel: J1908+0621 } 
    \label{fig:LHAASO}
\end{figure}

In Fig.~\ref{fig:LHAASO}, we illustrate the differential energy spectra corresponding to the sources J2226+6057 and J1908+0621. The data points were obtained experimentally~\cite{cao2021ultrahigh}, and the graphical representation includes the optimal power-law fit alongside the predicted flux under the assumption of subluminal LIV. By following these guidelines, we conduct a $\chi^2$ test and constrain the mass scale $E_{\rm LIV}$ at a $95 \%$ confidence level (CL). The results of the analysis are presented in the Table~\ref{tab:LHAASO}.
\begin{table}[h]
    \centering
  \begin{tabular}{|c|c|c|}
\hline 
Source & $L, \mbox{ kpc}$ & Bound  $E_{\rm LIV}$, $10^{20} \mbox{ GeV}$ \\
\hline 
Crab Nebula & 2 & $0.5$  \\ 
\hline 
J2226+6057 & 0.8 & $1.5$  \\ 
\hline 
J1908+0621 & 2.37 & $2.1$ \\ 
\hline 
\end{tabular} 
    \caption{The $95\%$ CL constraints on LIV mass scale from 3 sub-PeV sources observed by LHAASO. }
    \label{tab:LHAASO}
\end{table}

The simplest bound for the superluminal polarization came from the photon decay. Substituting LHAASO maximal energy $1.4$ PeV into eq.~(\ref{DecayBound}), one obtains
\begin{equation}
\label{SuperLumBound}
      E_{\rm LIV} > 2.6 \cdot 10^{24}\, \mbox{GeV}.
\end{equation}

The combined bound on $E_{\rm LIV}$ from super- and subluminal polarization is as follows. Supposing superluminal dispersion for these detected very-high-energy photons, one obtains the bound (\ref{SuperLumBound}). Supposing subluminal dispersion, we should take the best bound from Table~\ref{tab:LHAASO},  $E_{\rm LIV} > 2.1 \cdot 10^{20}\, \mbox{GeV}$. Not knowing the real initial polarization, we should take the worst bound from two  aforementioned ones, so we take finally the value
\begin{equation}
\label{LastConstraint}
E_{\rm LIV} > 2.1 \cdot 10^{20}\, \mbox{GeV}.
\end{equation}
for the constraint. No super/subluminal polarization here is  assumed.

\section{Discussion}
\label{sec:Discussion}

%
%
%
%
%

In this article we studied the suppression for the air shower formation for the general EFT with cubic dispersion relation for photons (Myers-Pospelov model). We test the hypothesis of LIV with a certain mass scale $E_{\rm LIV}$ with the photon spectra of the three LHAASO sources. Taking into account that these photons have not been completely vanished by the photon decay, and the spectra haven't been significantly depleted due to the suppression of air shower formation, we set the bound on LIV mass scale,
$$
E_{\rm LIV} > 2.1 \cdot 10^{20}\, \mbox{GeV}.
$$
Compared to the QED with quartic photon dispersion in which a model with certain LIV mass scale divided additionally to \enquote{super} and \enquote{subluminal} scenario~\cite{Rubtsov:2016bea}, we want to stress out that in cubic Myers-Pospelov model there are no additional scenarios. Super and subluminality is a property of a given photon polarization, not a whole model.  

Note that the obtained $E_{\rm LIV}$ bound is sufficiently trans-Planckian. Although it is hard to imagine any reasonable quantum gravity theory at these energy scales, trans-Plankian coupling constant still may appear due to fine tuning or feebly interaction between   QED sector and gravity.

\paragraph{Acknowledgements} The authors thank Jose Manuel Carmona, Tomislav Tersic, Gleb Suzdalov, Nickolay Martynenko, Pavel Demidov and Sergey Troitsky for helpful discussions. This work is supported by RSF foundation under contract 22-12-00253.

\bibliographystyle{ieeetr}
\bibliography{main}

\end{document}